\documentclass[a4paper,11pt]{article}
\usepackage{pos}

\title{Theory for the MUonE experiment}

\author*{A. Gurgone}

\onbehalf{on behalf of the MUonE collaboration}

\affiliation{Dipartimento di Fisica, Universit\`a di Pavia,\\ INFN, Sezione di Pavia,\\ Via A. Bassi 6, 27100 Pavia, Italy}

\emailAdd{andrea.gurgone01@ateneopv.it}

\abstract{
  The MUonE experiment aims at providing a new independent evaluation of the leading hadronic contribution to the muon anomalous magnetic moment.
  The proposed method is based on the measurement of the running of the QED coupling in the space-like momentum region, through the elastic scattering of high-energy muons on atomic electrons.
  In this proceeding, the status of the theoretical effort for the MUonE experiment is reviewed.
  In order to achieve a competitive determination, the differential cross section of muon-electron scattering must be computed with an unprecedented precision of 10~ppm, which requires at least next-to-next-leading order corrections and the resummation of large logarithms. 
  In addition, the simulation of the experiment needs an exhaustive theoretical description of the possible background processes, such as the lepton pair production in muon-nucleus scattering.
  All these processes must be implemented in a fully exclusive Monte Carlo event generator in order to be used in the MUonE data analysis.
}

\FullConference{The European Physical Society Conference on High Energy Physics (EPS-HEP2023)\\
 21-25 August 2023\\
Hamburg, Germany\\}

\begin{document}
\maketitle

\section{Introduction}

The discrepancy between the theoretical~\cite{Aoyama:2020ynm} and experimental~\cite{Muong-2:2023cdq} value of the muon anomalous moment $a_\mu = (g_\mu -2)/2$ is one of the most intriguing anomalies in the Standard Model. 
The main source of theoretical uncertainty is given by the evaluation of the Hadronic Vacuum Polarisation~(HVP) contribution $a_\mu^{\text{HVP}}$, which cannot be computed perturbatively. 
This quantity is usually determined though the time-like dispersive approach, which applies the optical theorem to the measurement of the hadronic production in $e^+e^-$ collisions.
However, the accuracy of this method is limited by the presence of several resonances in the low-energy cross section and by the combination of data from different experiments.
Furthermore, the results reported in~\cite{Aoyama:2020ynm} disagree with the BMW lattice prediction~\cite{Borsanyi:2020mff} and the CMD-3 measurement of the pion form factor~\cite{CMD-3:2023alj}.
These tensions suggest possible inaccuracies in the determination of $a_\mu^{\text{HVP}}$, inhibiting a reliable comparison between the theoretical and experimental value of the muon $g-2$.

A novel approach to evaluate $a_\mu^{\text{HVP}}$ has been proposed in~\cite{Calame:2015fva}. 
It is based the measurement of the QED coupling $\alpha(t)$ in the space-like region $t<0$, where the HVP contribution is a smooth function.
The running of the QED coupling can be written as
\begin{equation}
\alpha(t) = \frac{\alpha(0)}{1-\Delta\alpha(t)}\,, \quad\;\; \Delta\alpha(t) = \Delta\alpha_{\text{lep}}(t) + \Delta\alpha_{\text{had}}(t)\,,
\end{equation}
where $\alpha(0)=\alpha$ is the fine-structure constant. 
The hadronic contribution $\Delta\alpha_{\text{had}}(t)$ can be extracted by subtracting the purely leptonic part $\Delta\alpha_{\text{lep}}(t)$, which can be computed in perturbative QED, from the measurement of $\Delta\alpha(t)$.
At this point, the leading HVP contribution $a_\mu^{\text{HVP, LO}}$ can be evaluated from $\Delta\alpha_{\text{had}}(t)$ by using the integral relation~\cite{Lautrup:1971jf}
\begin{equation}
a_\mu^{\text{HVP, LO}} = \frac{\alpha}{\pi} \int^1_0 \text{d}x (1-x) \Delta \alpha_{\text{had}} \left[t(x)\right]\,, \quad\;\;
t(x) = - \frac{x^2m_\mu^2}{1-x} < 0 \,.
\end{equation}
The function $\Delta\alpha(t)$ can be obtained from the cross section of a $t$-channel process, such as the elastic muon-electron scattering~\cite{Abbiendi:2016xup}.
In this regard, the MUonE experiment has been proposed to the the CERN SPS Committee~\cite{MUonE:LoI}.
The experimental concept is based on the elastic scattering of a muon or antimuon beam of 160 GeV on atomic electrons bound in a low-$Z$ target.
The MUonE experiment will provide an independent evaluation of $a_\mu^{\text{HVP, LO}}$ with an uncertainty below the percent level, comparable with the results from the time-like dispersive approach and the lattice simulations.
The prospect of evaluating the HVP contribution beyond the leading order~\cite{Balzani:2021del} or using alternative methods~\cite{Ignatov:2023wma} is currently under study.
The possibility of observing new physics effects in the muon-electron elastic scattering has been investigated and excluded in~\cite{Masiero:2020vxk}.

\section{Muon-electron scattering}

The precision goal of the MUonE experiment requires a theoretical computation of the muon-electron elastic scattering with an accuracy of 10~ppm on the differential cross section~\cite{Banerjee:2020tdt}.
In order to achieve this level of precision, the calculation must include QED corrections at least at the Next-to-Next-to-Leading Order~(NNLO) accuracy~\cite{Alacevich:2018vez,CarloniCalame:2020yoz,Budassi:2021twh,Broggio:2022htr,Mastrolia:2017pfy,DiVita:2018nnh,Bonciani:2021okt,Fael:2022rgm,Ahmed:2023htp,Badger:2023xtl}, as well as electroweak~\cite{Alacevich:2018vez} and hadronic~\cite{Fael:2018dmz,Fael:2019nsf} contributions.
Since the electron mass $m_e$ cannot be neglected, a significant challenge is given by the inclusion of the massive terms in the two-loop diagrams.
This can be achieved by using the massification technique~\cite{Engel:2018fsb} or dispersion relations~\cite{Budassi:2021twh,Awramik:2006uz}.  

The ${\mu^\pm e^- \to \mu^\pm e^-}$ scattering, including all radiative corrections, must be implemented in a fully differential Monte Carlo framework in order to be used in the MUonE data analysis.
In this regard, two independent codes, \textsc{Mesmer}~\cite{Alacevich:2018vez,CarloniCalame:2020yoz,Budassi:2021twh} and \textsc{McMule}~\cite{Banerjee:2020rww,Broggio:2022htr}, are under development.
In addition, \textsc{Mesmer} includes the simulation of ${\mu^\pm e^- \to \mu^\pm e^- \ell^+\ell^-}$ with~$\ell=\{e,\hspace{1pt}\mu\}$ and ${\mu^\pm e^- \to \mu^\pm e^- \pi^0}$, which are possible sources of background~\cite{Budassi:2021twh,Budassi:2022kqs}.

\begin{figure}[t]
\centering
\includegraphics[width=.48\textwidth]{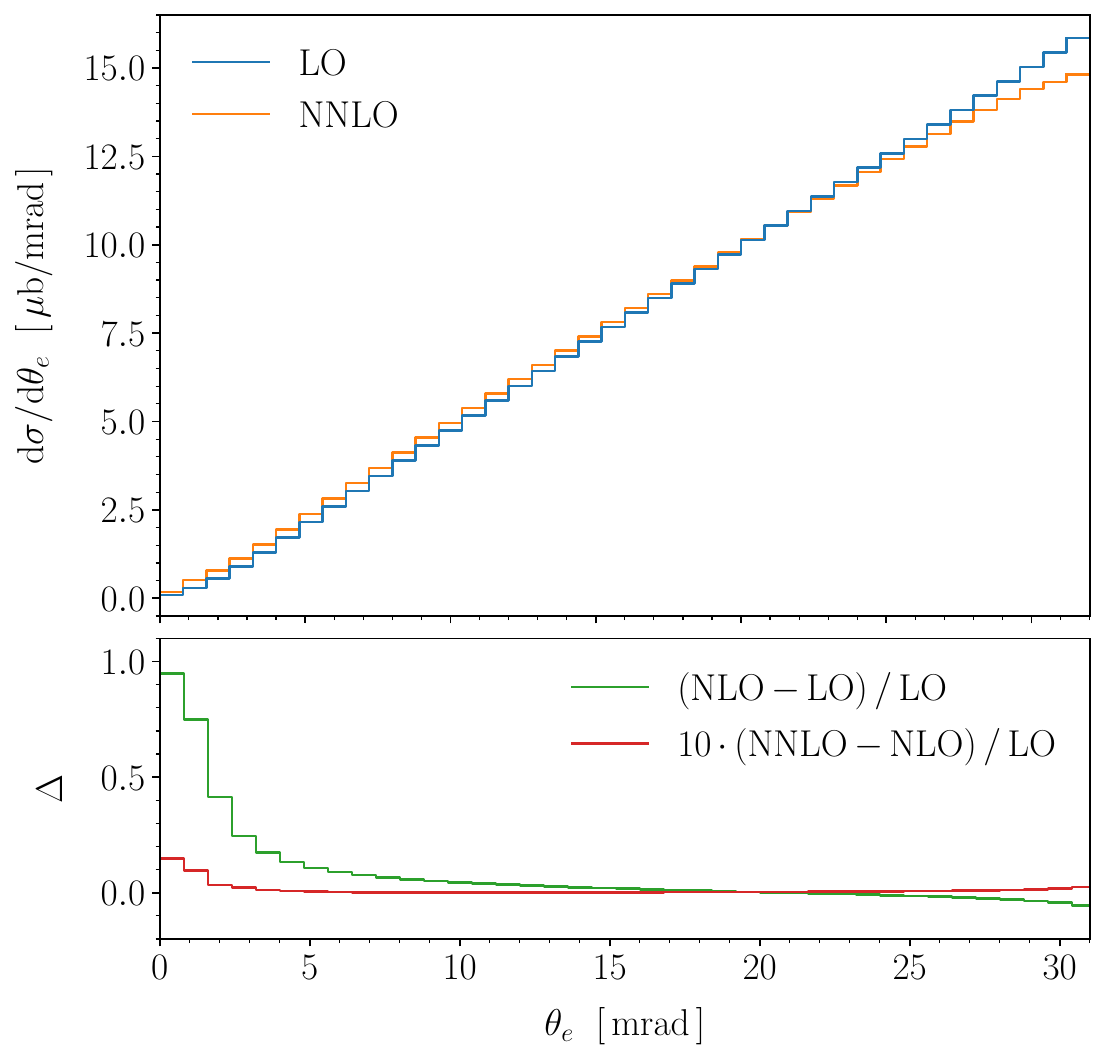}
\hspace{.025\textwidth}
\includegraphics[width=.48\textwidth]{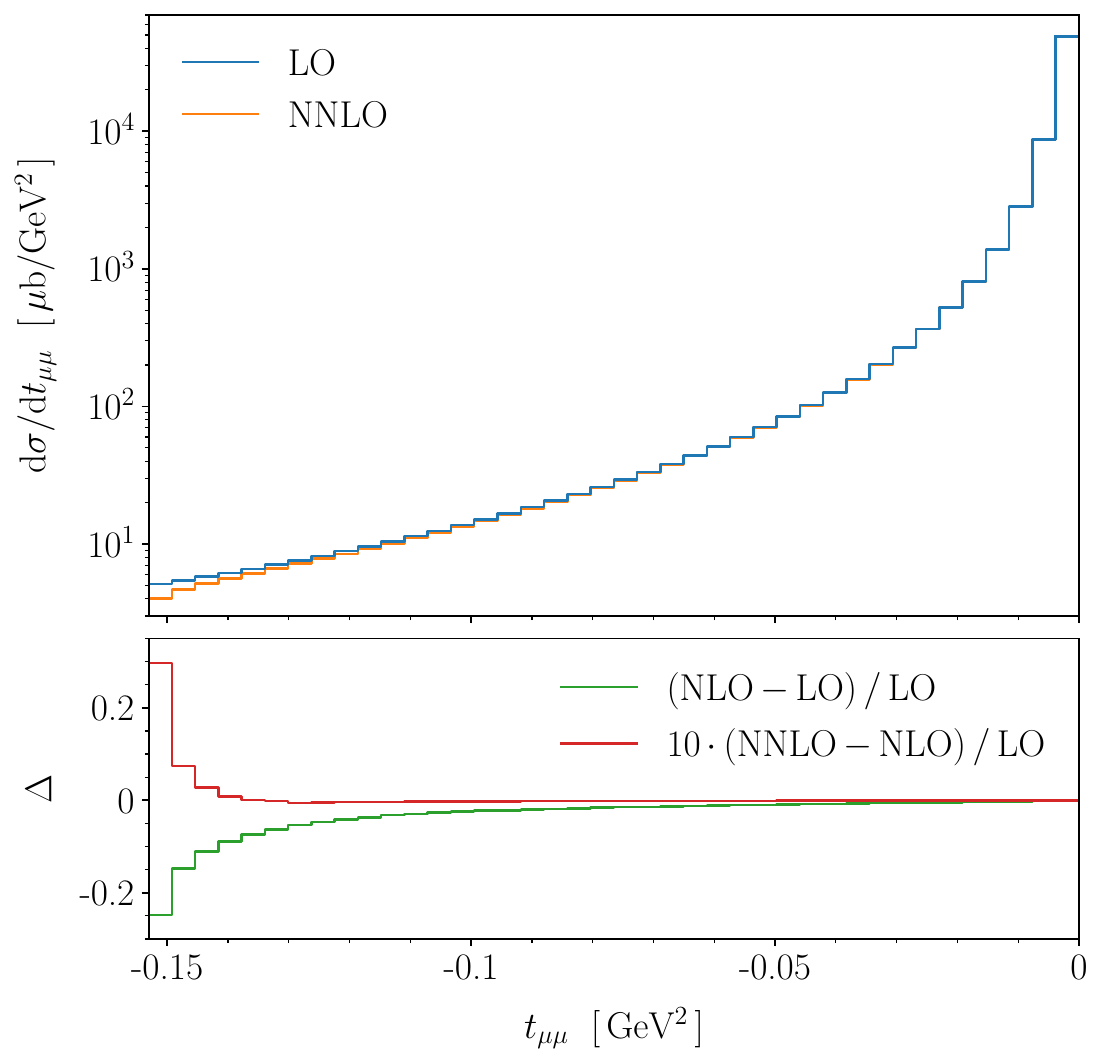}
\caption{\label{fig:sig} Differential cross sections of $\mu^+ e^- \to \mu^+ e^-$ with respect to the electron angle (left panel) and the transferred space-like momentum (right panel), obtained with \textsc{Mesmer}. The relative impact of the radiative corrections is reported in the two lower panels. 
}
\end{figure}

The \textsc{Mesmer} result for the distribution of the electron angle in the laboratory frame and the transferred momentum between the initial and final muon is reported in Figure~\ref{fig:sig}.
Both distributions include exact photonic and leptonic QED corrections up to NNLO, with the only exception of two-loop box diagrams, which are approximated through an approach based on the Yennie–Frautschi–Suura~(YFS) formalism~\cite{Yennie:1961ad}.
In order to reproduce the MUonE acceptance, the following cuts are applied: ${E_\mu > 10.23}$~GeV, ${E_e > 200}$~MeV, ${\theta_\mu < 4.84}$~mrad, and ${\theta_e < 32}$ mrad.
An additional cut on the minimum muon angle, namely ${\theta_\mu > 0.2}$~mrad, is imposed to suppress the background from lepton pair production, which is expected to be larger in this region. 

However, the radiative corrections for the MUonE experiment are enhanced by large logarithms, which reduce the accuracy of the fixed-order expansion. 
A first class of logarithms is due to the presence of widely different kinematic scales, specifically the electron and muon mass, the centre-of-mass energy and the transferred momentum $t$. 
Additional large logarithms arise from the cuts on the emission of real radiation, introduced to select the elastic events. 
Hence, the precision goal of 10~ppm requires the resummation of the logarithmic terms with a leading or next-to-leading accuracy.
This can be achieved through the implementation of a QED parton shower matched to the fixed-order corrections, as done in \textsc{BabaYaga}~\cite{Balossini:2006wc}. 
An alternative approach is to exploit the YFS exponentiation, as done in \textsc{Sherpa}~\cite{Schonherr:2008av}.

\section{Muon-nucleus scattering}

Since in the MUonE experiment the initial-state electrons are bound in a low-$Z$ atomic target, the muon-nucleus scattering is expected to be the main source of experimental background~\cite{Abbiendi:2021xsh}.
The lepton pair production ${\mu^\pm X \to \mu^\pm X\hspace{1pt} \ell^+ \ell^-}$ with $\ell=\{e,\hspace{1pt}\mu\}$ is a particularly important process, because it can resemble an elastic event if one of the final leptons is not detected.
For this reason, a new fully differential computation of the process has been included in \textsc{Mesmer}~\cite{Abbiendi:2024swt}, improving the existing \textsc{Geant4} implementation~\cite{Bogdanov:2006kr}, in which the muon deflection is neglected.

\begin{figure}[t]
\centering
\includegraphics[width=.825\textwidth]{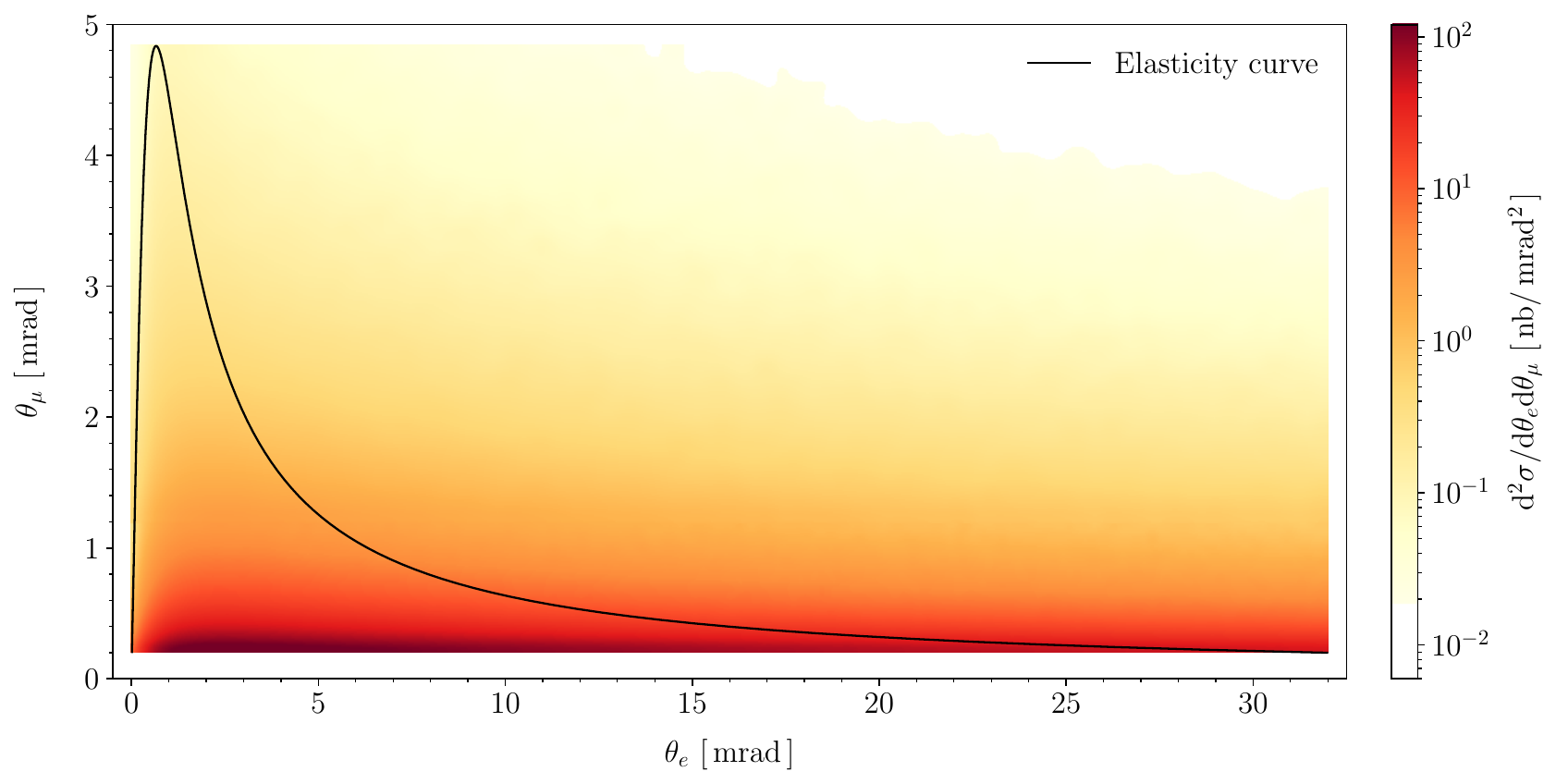}
\caption{\label{fig:bgr} Differential cross section of ${\mu^+ \hspace{1pt} C \to \mu^+ \hspace{1pt} C \hspace{1pt} e^+ e^-}$ in the $\theta_\mu \hspace{-2pt} - \hspace{-1pt} \theta_e$ plane for events with only one electron in the detector acceptance. The elasticity curve corresponding to $\mu^+ e^- \to \mu^+ e^-$ events is shown.
}
\end{figure}

The process is described as a scattering of a muon in an external Coulomb field with the addition of a form factor to describe the nuclear charge distribution.
The lepton pair can be emitted either from the incident muon through photon radiation or from the virtual photon exchanged with the nuclear field.
The theoretical uncertainty is estimated by using different models of form factors, following the same approach as in~\cite{Heeck:2021adh}.
The result for ${\mu^+ \hspace{1pt} C \to \mu^+ \hspace{1pt} C \hspace{1pt} e^+ e^-}$ in the $\theta_\mu \hspace{-1pt} - \theta_e$ plane is reported in Figure~\ref{fig:bgr}, including only events with one electron in the detector acceptance.
The cross section is strongly enhanced for small muon angles, justifying the cut on the minimum value.

\section{Conclusions}

The MUonE experiment will provide a new independent determination of $a_\mu^{\text{HVP}}$, which will contribute to shed light on the muon $g\hspace{-2pt}-\hspace{-2pt}2$ anomaly.
The signal extraction requires a theoretical calculation of elastic muon-electron scattering with an accuracy of 10 ppm on the differential cross section.
In order to achieve this challenging precision, the computation must include at least NNLO corrections with $m_e\neq 0$, as well as the resummation of large logarithms.
The experimental simulation also requires a detailed theoretical description of the possible background processes, such as the lepton pair production in muon-nucleus scattering.
In this regard, the development of two independent Monte Carlo codes, \textsc{Mesmer} and \textsc{McMule}, is currently in progress.

\acknowledgments

The author is grateful to all members of the MUonE project for their valuable collaboration.

\bibliographystyle{JHEP}
\bibliography{biblio.bib}

\end{document}